\documentclass[a4paper,11pt]{article}
\usepackage{pos}
\newcommand{\sh}[1]{#1\hskip -6pt  / }

\title{Non Nucleonic Components in Short Nuclear Distances}

\author*{Misak Sargsian}

\affiliation{Florida International University, 
   Miami, FL  33199, USA}

\emailAdd{sargsian@fiu.edu}

\abstract{One of the important features of nuclear forces is their strong repulsive nature 
at short ($\le 0.5-0.6$~Fm) distances which prevents atomic nuclei from collapsing, thus
guarantying the stability for the visible matter.  However the dynamical nature of this 
repulsion (referred to as a nuclear core) is as elusive as ever.  We present the 
study of nuclear dynamics at extremely large internal 
momenta in the deuteron dominated by the nuclear core. It is demonstrated that the paradigm shift 
in the description of the deuteron consisting of proton and neutron to the description of 
the deuteron  as a pseudo-vector composite system in which proton and neutron is 
observed in high energy  electro-disintegration processes results in the emergence of 
a new structure. We demonstrate that this new structure  can exist only if it emerges from 
pre-existing non-nucleonic component in the deuteron.  The study of the dynamics of the
predicted new structure is presented focusing on the question if it allows to understand 
the anomaly observed in the recent experiment at Jefferson  Lab that probed deuteron 
structure at internal momenta above 800 MeV/c.
}

\FullConference{The XVIth Quark Confinement and the Hadron Spectrum Conference (QCHSC24)\\
 19-24 August, 2024\\
 Cairns Convention Centre, Cairns, Queensland, Australia\\}

\begin{document}
\maketitle

\section{Introduction}
Understanding  the dynamics of 
the transition between hadronic to quark-gluon phases  is one of the outstanding issues in  strong 
interaction physics. 
For cold dense nuclear matter such transitions are relevant for  the dynamics 
that can exist  at the cores of neutron stars and can set 
the limits for the  matter density before it collapses to  the black hole.
There are few  options to investigate such transitions in terrestrial experiments 
These include  studying nuclear medium modification of quark-gluon structure of  bound nucleons 
in semi-inclusive processes which allow to control inter-nucleon distances\cite{Melnitchouk:1996vp} 
or  probing deep inelastic scattering from nuclear target at Bjorken $x>1$\cite{Sargsian:2007gd,Fomin:2010ei,Freese:2015ebu}.

\begin{figure}[h]
\begin{center}
\includegraphics[width=9.2cm,height=5.2cm]{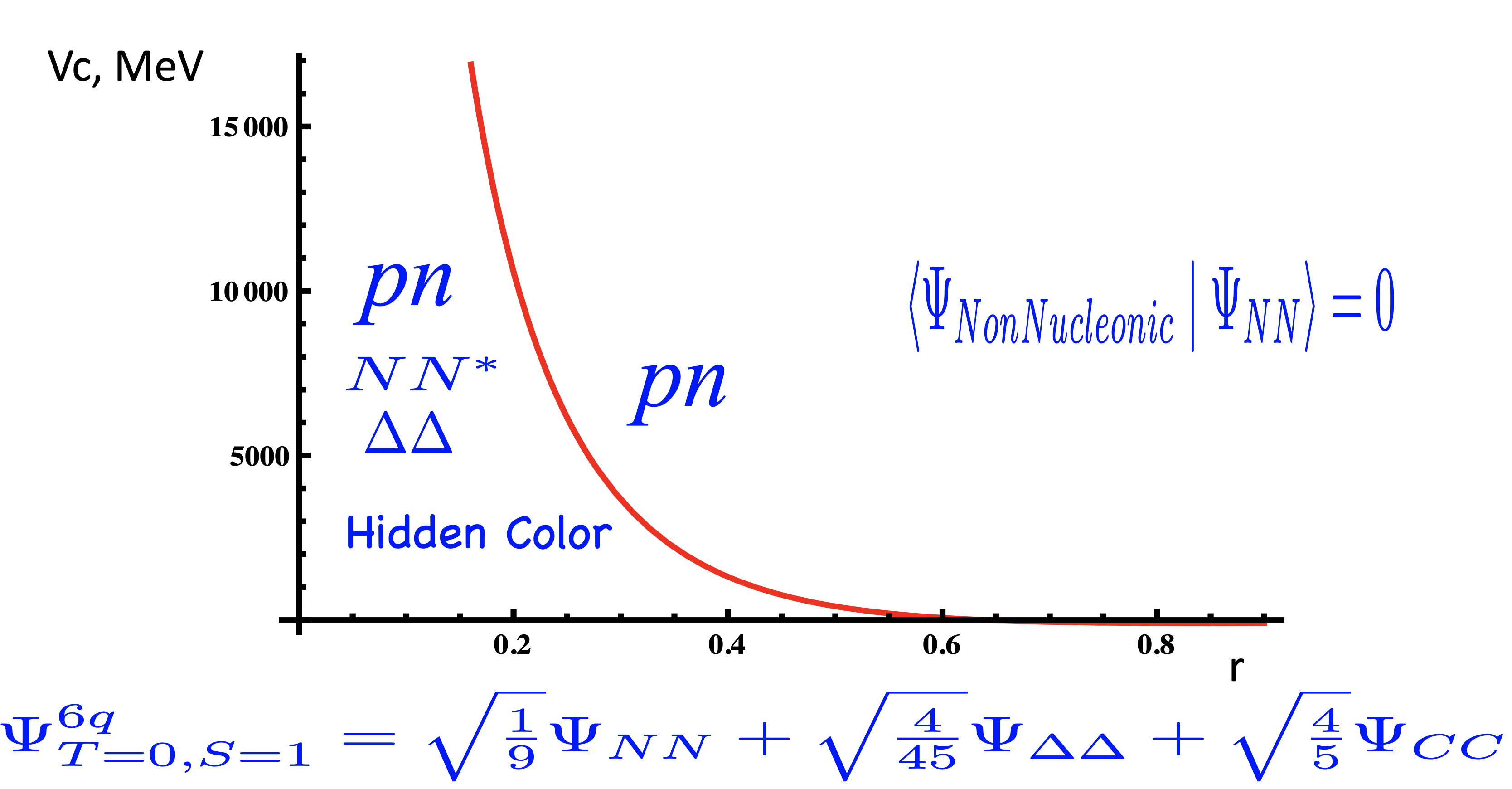}
\vspace{-0.5cm}
\end{center}
\caption{Model for the NN repulsive core with substantial non-nucleonic component.}
\label{NNcore}
\end{figure}

Another venue in exploration of non-nucleonic components in nuclei is the probing the nuclear repulsive core.  The nuclear 
repulsive core is a unique property of nuclear forces that keeps nuclei from collapsing and provides condition for 
the nuclear density to saturate. Its dynamics is still elusive, however QCD gives a new perspective on the dynamical origin of 
the nuclear core. In the $NN$ system at very short distances  the  QCD predicts the existence of substantial component 
due to non-nucleonic $\Delta\Delta$ as well as hidden color components\cite{Harvey:1980rva,Brodsky:1985gt}  that contribute
almost 90\% of the strength  at distances dominated by the repulsion (Fig.\ref{NNcore}).  
In such a scenario the repulsion is due to the orthogonality between the wave functions of the observed NN state and 
the non-nucleonic component dominating in the core. 

To probe the validity of such a scenario for the generation of nuclear repulsion one  needs to probe deuteron wave function at internal 
momenta $\gtrsim 800$~MeV/C.  In the present work, new approach\cite{Sargsian:2022rmq} is suggested in probing 
the deuteron at extremely large internal momenta.

\section{Deuteron on the light front~(LF)}
Non-relativistic picture of the deuteron suggests that 
the observations of total isospin, $I=0$,  total spin, $J=1$  and positive parity, $P$,  together with the  relation,
$P= (-1)^l$,  indicate  that the deuteron consists of bound proton and neutron in S- and D- partial wave states.

However, for the deuteron structure with internal momenta comparable to  the nucleon rest mass 
the nonrelativistic framework is not valid requiring a consistent account for the relativistic effects.
There are  several theoretical approaches for  accounting for  relativistic effects in the deuteron wave function 
(see e.g.  Refs.\cite{Frankfurt:1977vc,Buck:1979ff,Arnold:1980zj,Dymarz:1986km,Carbonell:1995yi}). 
In our approach the  relativistic effects are accounted for similar to the one used in QCD (see e.g. \cite{Feynman:1973xc,Brodsky:1997de})
for  calculation of  quark distribution  in  hadrons, in which light-front (LF) description of 
the scattering process allows to suppress vacuum fluctuations that overshadow  the composite structure of the hadron. 
 Here one needs to identify the process in which 
the deuteron structure is probed. For this  we consider high $Q^2$  electrodisintegration process:
\vspace{-0.2cm}
\begin{equation}
e + d \rightarrow e^\prime + p + n,
\label{reaction}
\vspace{-0.2cm}
\end{equation}
in which one of the nucleons are struck by the incoming probe and the spectator nucleon is probed with momenta comparable 
to the nucleon mass.  If one can neglect (or remove) the effects related to final state interactions of two outgoing nucleons, 
then the above reaction at high $Q^2$, measures the probability of observing  proton and neutron in the deuteron 
with large relative  momenta.  
In such a formulation the deuteron is not a composite system consisting of a proton and neutron, but it is 
a composite pseudo - vector ($J=1$, $P=+$)  ``particle" from which one extracts a proton and neutron.
Thus we formulate the question not as how to describe relativistic motion of proton and neutron in the deuteron,  but
 how such a  proton and neutron are produced at   such   extreme conditions relating it  
to the dynamical  structure of the LF deuteron wave function. 
In such formulation the latter  may include internal elastic $pn \rightarrow pn$ as well as inelastic 
$\Delta\Delta\to pn$, $N^*N\to pn$ or $N_cN_c\to pn$ transitions. Here, $\Delta$ and $N^*$ denote $\Delta$-isobar and $N^*$ resonances,  while 
$N_c$ is a  color octet baryonic state contributing to  the hidden-color component in the deuteron.\\
The framework for calculation of reaction (\ref{reaction}) in the  relativistic domain  is the LF
approach\cite{Sargsian:2022rmq}  in which  one introduces  the LF deuteron wave function:
\begin{equation}
\psi_{d}^{\lambda_d}(\alpha_i,p_\perp,\lambda_1\lambda_2) = - {\bar u(p_2,\lambda_2)\bar u(p_1,\lambda_1) \Gamma^\mu_{d} \chi_\mu^{\lambda_d}\over 
 {1 \over 2} ( m_d^2 - 4 {m_N^2 + p_\perp^2\over \alpha_i(2-\alpha_i)})\sqrt{2(2\pi)^3}} = 
 = -\sum\limits_{\lambda_1^\prime}\bar u(p_1,\lambda_1) 
\Gamma^\mu_{d}\gamma_5 {\epsilon_{\lambda_1,\lambda_1^\prime}\over \sqrt{2}}u(p_1,\lambda^\prime_1),
\vspace{-0.3cm}
 \label{dwave_lf}
\end{equation}
where $\alpha_i = 2{p_{i+}\over p_{d+}}$, ($i=1,2$) are LF momentum fractions of proton and neutron, outgoing from 
the deuteron with $\alpha_1+\alpha_2 = 2$ and in the second part we absorbed the 
propagator into the vertex function and used crossing symmetry.
Here  $u(p,\lambda)$'s are the LF bi-spinors of the proton and neutron and 
$\epsilon_{i,j}$ is the two dimensional Levi-Civita tensor, with $i,j=\pm 1$ nucleon helicity. Since the deuteron is a pseudo-vector 
``particle",  due to $\gamma_5$ in Eq.(\ref{dwave_lf}), the vertex $\Gamma^\mu_d$ is a four-vector which we can construct
in a general form that explicitly satisfies time reversal, parity and charge conjugate symmetries. Noticing  that  
at the   $ d\to pn$ vertex  on the light-front 
the  "-"  ($p^- = E-p_z$) components 
of the four-momenta of the particles are not conserved, in addition to the four-momenta of two nucleons, $p_1^\mu$ and $p_2^\nu$,  
one  has an additional four-momentum:
\vspace{-0.2cm}
\begin{equation}
\Delta^\mu \equiv p_1^\mu + p_2^\mu - p_d^\mu   \equiv (\Delta^-, \Delta^+, \Delta_{\perp}) = (\Delta^-, 0, 0),
\label{DeltaDefLF}
\end{equation}
where
\vspace{-0.4cm}
\begin{eqnarray}
\vspace{-0.4cm}
& & \hspace{-0.8cm} \Delta^- =  p_1^- + p_2^- - p_d^- 
= 
{4\over p_d^+}\left[m_N^2 - {M_d^2\over 4} + k^2\right] ; k = \sqrt{{m_N^2 + k_\perp^2\over \alpha_1(2-\alpha_1)} - m_N^2} ;  \ \ \ \alpha_1 = {E_k + k_z\over E_k},
\vspace{-0.6cm}
\label{Delta-}
\end{eqnarray}
with  $E_k = {m^2 + k^2}$.
With  $p_1^\mu$, $p_2^\mu$ and $\Delta^\mu$ 4-vectors the $\Gamma_d^\mu$  is constructed in the form:
\vspace{-0.2cm}
\begin{eqnarray}
 & & \hspace{-0.2cm} \Gamma_d^{\mu}= \Gamma_{1}  \gamma^{\mu} +\Gamma_{2} 
 {{(p_1-p_2)^{\mu}}\over {2m_N}} + \Gamma_{3} 
 {{\Delta^{\mu}}\over {2m_N}}+
 \Gamma_{4} 
 {{(p_1-p_2)^{\mu} \sh{\Delta}}\over { 4m_N^2}}     \nonumber \\
& &    +  i \Gamma_{5} \frac{1}{4 m_N^{3}} 
\gamma_{5} \epsilon^{\mu \nu \rho \gamma}(p_{d})_\nu (p_1-p_2)_{\rho} (\Delta)_\gamma 
 +   \Gamma_{6} \frac{ \Delta^{\mu} \sh{\Delta}}{4m_N^{2}},
\label{vertex}
\end{eqnarray}where $\Gamma_i$,($i=1,6$) are  scalar functions. 
(see also Refs\cite{Carbonell:1995yi}).

\section{High energy approximation} 
For the large $Q^2$ limit, the LF momenta for reaction (\ref{reaction})  are chosen as follows:
 \vspace{-0.2cm}
 \begin{eqnarray}
&&\hspace{-0.2cm}p_d^\mu  \equiv  (p_{d}^-, p_{d}^+ ,p_{d\perp}) =  \left({Q^2\over x\sqrt{s}}\left[1 + {x\over \tau} - \sqrt{1+ {x^2\over \tau}}\right] \right., 
 \left.                          {Q^2\over x\sqrt{s}} \left[1 + {x\over \tau} +  \sqrt{1+ {x^2\over \tau}}\right], 0_\perp\right)    \nonumber \\
&&q^\mu \equiv  (q^{-},q^{+}, q_{\perp}) =  \left({Q^2\over x\sqrt{s}}\left[1 - x + \sqrt{1+ {x^2\over \tau}}\right], \right.
\left.         {Q^2\over x\sqrt{s}}\left[1 - x - \sqrt{1+ {x^2\over \tau}}\right], 0_\perp\right),  
\label{refframeQ}
\end{eqnarray}
where $s = (q+p_d)^2$, $\tau={Q^2\over M_d^2}$ and $x = {Q^2\over M_dq_0}$, with $q_0$ being the virtual photon energy in 
the deuteron rest frame.  The high energy nature of this process results in,  $p_d^+ \sim \sqrt{Q^2}\gg m_N$, which makes 
$\Delta^-$ term to be suppressed by the large $p_d^+$ factor in  Eq.(\ref{Delta-}), allowing to treat ${\Delta^-\over 2m_N}$  
as a small parameter.

Keeping the leading, ${\cal O}^0({\Delta^-\over 2m_N})$, terms in Eq.(\ref{vertex})  and using the boost invariance of the wave function 
we calculate it in the CM of the deuteron\cite{Sargsian:2022rmq} to  obtain:
 \vspace{-0.2cm}
 \begin{eqnarray}
& & \hspace{-0.9cm} \psi_{d}^{\lambda_d}(\alpha_i,k_\perp)  \hspace{-0.1cm} =    -\hspace{-0.5cm} \sum\limits_{\lambda_2,\lambda_1,\lambda_1^\prime}
\hspace{-0.3cm}\bar u(-k,\lambda_2) 
 \hspace{-0.1cm}\left\{ \Gamma_1\gamma^\mu \hspace{-0.1cm}+ \hspace{-0.1cm}
\Gamma_2{{\tilde k}^\mu\over m_N}  + \right.    
\hspace{-0.3cm}  \left.    \sum\limits_{i=1}^{2}
 i\Gamma_5{1\over 8m^3_N}\epsilon^{\mu + i  -}p^{\prime +}_{d} k_{i} \Delta^{\prime-}\right\}   
  \gamma_5 \hspace{-0.1cm}{\epsilon_{\lambda_1,\lambda_i^\prime}\over \sqrt{2}} u(k,\lambda^\prime_1)s_\mu^{\lambda_d},   
\label{dwave_lf4}
\end{eqnarray}
where $\tilde k^\mu = (0,k_z,k_\perp)$  with $k_\perp = p_{1\perp}$,
$k^2 = k_z^2 + k^2_\perp$ and  $E_{k} = 
{\sqrt{S_{NN}}\over 2}$  and  $s_\mu^{\lambda_d} = (0,{\bf s^\lambda_d})$, with
$s_d^1 = - {1\over \sqrt{2}} (1,i,0)$,   $s_d^1 = {1\over \sqrt{2}} (1,-i,0)$,   $s_d^0 = (0,0,1)$
and
$p^{\prime +}_{d}   = \sqrt{s_{NN}}, \    
\Delta^{\prime-}  = {1\over \sqrt{s_{NN}}}\left[ {4(m_N^2 + k_\perp^2)\over\alpha_1(2-\alpha_1)}-M_d^2\right]$.
Since the term related to  
$\Gamma_5$ is proportional to  ${4(m_N^2 + k_\perp^2)\over\alpha_1(2-\alpha_1)}-M_d^2$, which 
diminishes at small momenta,  only the $\Gamma_1$ and $\Gamma_2$ terms will contribute in the  nonrelativistic limit  defining 
the  $S$- and $D$- components of the deuteron.  Thus, the LF wave function in Eq.(\ref{dwave_lf4}) provides a smooth transition to 
the non-relativistic deuteron wave function.
This can be seen by expressing Eq.(\ref{dwave_lf4})  through two-component spinors:
\vspace{-0.2cm}
\begin{eqnarray}
 \psi_d^{\lambda_d}(\alpha_1,k_t,\lambda_1,\lambda_2)  =  
\sum\limits_{\lambda_1^\prime}\phi^\dagger_{\lambda_2} \sqrt{E_k}\left[  {U(k)\over \sqrt{4\pi}} {\bf \sigma s_d^{\lambda_d}}\right. - 
  - \left.    {W(k)\over \sqrt{4\pi}  \sqrt{2}}\left( { 3{\bf (\sigma k)(k s_d^\lambda)}\over k^2} - {\bf \sigma s_d^\lambda} \right) +\right. & &    \nonumber \\  
  \left.    
 (-1)^{1+\lambda_d\over 2} P(k)Y_{1}^{\lambda_d}(\theta,\phi)\delta^{1,\mid \lambda_d\mid}
  \right]
{\epsilon_{\lambda_1,\lambda_1^\prime}\over \sqrt{2}} \phi_{\lambda^\prime_1}. & &
\label{WF_LF}
\end{eqnarray}
Here the first two terms have explicit $S$- and $D$- structures  where the radial functions are defined as: 
\begin{eqnarray}
& \hspace{-0.4cm} U(k)  = & {2\sqrt{4\pi} \sqrt{E_k}\over 3}\left[\Gamma_1(2+{m_N\over E_k}) + \Gamma_2{k^2\over m_N E_k}\right]\nonumber \\
& \hspace{-0.4cm} W(k)  = &  {2\sqrt{4\pi} \sqrt{2E_k}\over 3}\left[\Gamma_1(1-{m_N\over E_k}) -   \Gamma_2{k^2\over m_N E_k}\right].  
\label{radialwaves}
\end{eqnarray}
This relation is  known for   $pn$-component deuteron wave function\cite{Frankfurt:1981mk,Carbonell:1995yi}, which allows us to 
model the LF wave function through known radial $S$- and $D$- wave functions    evaluated at LF relative momentum $k$ defined  in
Eq.(\ref{Delta-}).

The new result is that due to the $\Gamma_5$ term  there is an additional leading contribution, 
which because of the relation  $Y^{\pm}_1(\theta,\phi) = \mp i\sqrt{3\over 4\pi}\sum\limits_{i=1}^{2}{ (k\times s_d^{\pm 1})_z\over k}$, has a 
$P$-wave like structure, where the $P$- radial function  is defined as:
\begin{eqnarray}
&  \hspace{-0.4cm}P(k)  = &  \sqrt{4\pi}  {\Gamma_5(k) \sqrt{E_k}\over \sqrt{3}}{k^3\over m_N^3}.   
\label{Pradialwave}
\end{eqnarray}
 
The unusual feature of our  result is that the $P$-wave is ``incomplete", that is it  contributes only 
for $\lambda_d = \pm 1$ polarizations of the deuteron.
 
\section{Light front density matrix of the deuteron}
Defining deuteron LF momentum distribution $n_d(k,k_\perp)$ and
density matrix: 
\vspace{-0.2cm}
\begin{equation}
\vspace{-0.2cm}
n_d(k,k_\perp)    =   {1\over 3}\sum\limits_{\lambda_d=-1}^{1}\mid \psi_d^{\lambda_d}(\alpha,k_\perp)\mid^2  \ \ \ 
\mbox{and} \ \ \ \rho_d({\alpha,k_\perp}) = {n_d(k,k_\perp)\over 2-\alpha},
\label{momdist1}
\end{equation}
one obtains
\vspace{-0.2cm}
\begin{equation}
\vspace{-0.2cm}
n_d(k,k_\perp)    =   {1\over 3}\sum\limits_{\lambda_d=-1}^{1}\mid \psi_d^{\lambda_d}(\alpha,k_\perp)\mid^2 = 
=  {1\over 4\pi} \left( U(k)^2 +  W(k)^2 + {k_\perp^2\over k^2} P^2(k)\right)
\label{momdist2}
\end{equation}
with 
$\int \rho_d({\alpha,k_\perp}) {d\alpha\over \alpha} = 1$, $\int\alpha \rho_d({\alpha,k_\perp}) {d\alpha\over \alpha} = 1$ and 
$\int \left(U(k)^2 +  W(k)^2 + {2\over 3} P^2(k)\right)k^2 dk = 1$.
Due to the incompleteness of the $P$-wave structure our result predicts that LF momentum distribution for  
deuteron  depends  explicitly on the transverse component of the relative momentum on the light front. This is highly unusual 
result, implication of which will be discussed in the next section.

For polarized deuteron the quantity that can be probed in the reaction~(\ref{reaction}) the tensor asymmetry which we define as:
\begin{equation}
A_T = {n_d^{\lambda_d = 1}(k,k_\perp) +  n_d^{\lambda_d = -1}(k,k_\perp) - 2  n_d^{\lambda_d = 0}(k,k_\perp)\over  n_d(k, k_\perp)}.
\label{t20}
\end{equation}
Here because of the same incompleteness of the $"P-wave"$ structure one may expect more sensitivity that for unpolarized momentum 
distribution. 

\section{The new term and the non-nucleonic components in the deuteron:}
One of our main predictions is that the LF momentum distribution, Eq.(\ref{momdist2}) will explicitly depend on 
the transverse component of  the  deuteron internal momentum on the light front. 
Such  a dependence  is impossible for non-relativistic quantum mechanics of the 
deuteron since in this case the potential of the interaction is real (no inelasticities) and the solution of Lippmann-Schwinger equation 
for partial S- and D-waves satisfies the ``angular condition", according to which the momentum distribution in the unpolarized deuteron depends 
on the magnitude of the relative momentum only.  

In the relativistic 
domain the definition of the interaction  potential is not straightforward 
to claim  that the momentum distribution in Eq.(\ref{momdist2}) should satisfy the angular condition also in the relativistic case
(i.e. to be dependent only on the  magnitude of $k$). 

To check the situation in relativistic case one considers Weinberg type 
equation\cite{Weinberg:1966jm} on the light-front for NN scattering amplitudes, in which only nucleonic degrees are considered,
in the CM of the NN system.  One obtains\cite{Frois:1991wc}:
\begin{eqnarray}
&&T_{NN}(\alpha_i,k_{i\perp},\alpha_f,k_{f,\perp}) \equiv T_{NN}(k_{i,z},k_{i\perp},k_{f,z},k_{f,\perp}) =  V(k_{i,z},k_{i\perp},k_{f,z},k_{f,\perp}) \nonumber \\
& &  + \int V(k_{i,z},k_{i\perp},k_{m,z},k_{m,\perp}) 
 {d^3 k_m\over (2\pi)^3 \sqrt{m^2 + k_m^2}}{T_{NN}(k_{m,z},k_{m\perp},k_{f,z},k_{f,\perp})\over 4(k_m^2 - k_f^2)},
\label{TNN}
\end{eqnarray} 
where ``i", ``m" and ``f" subscripts correspond to initial, intermediate and final $NN$ states, respectively, and momenta $k_{i,m,f}$ 
are defined similar to Eq.(\ref{Delta-}). 

The realization of the angular condition for the relativistic case  requires:
 \begin{equation}
V(k_{i,z},k_{i\perp},k_{m,z},k_{m,\perp}) = V(\vec k_i^2, (\vec k_m-\vec k_i)^2),
\label{Vangcond}
\end{equation}
resulting in:
\begin{equation}
T_{NN}^{on \ shell}(k_{i,z},k_{i\perp},k_{m,z},k_{m,\perp}) = T^{on \ shell}_{NN}(\vec k_i^2, (\vec k_m-\vec k_i)^2)
\label{Tangcond}
\end{equation}
and the existence of the Born term in Eq.(\ref{TNN}) indicates that  the potential $V$ satisfies 
the same condition  in the  on-shell limit.   

For the off-shell potential\cite{Frois:1991wc} 
that requirements for  the potential $V$ to satisfy
angular condition in the on-shell limit and 
that it  can be constructed through the series of elastic $pn$ scatterings  result
to the   $V$ and $T_{NN}$ functions satisfying the similar  angular conditions (Eqs.(\ref{Vangcond},\ref{Tangcond})).
Using such a potential to calculate the LF deuteron wave function will result in a momentum distribution 
dependent only  on  the magnitude of the relative $pn$ momentum. 

Inclusion of the inelastic transitions will completely change the LF equation for the $pn$ scattering. 
For  example, the contribution of  $N^*N$ transition to the elastic $NN$ scattering:
\vspace{-0.3cm}
\begin{eqnarray}
&&T_{NN}(k_{i,z},k_{i\perp},k_{f,z},k_{f,\perp}) =  \int V_{NN^*}(k_{i,z},k_{i\perp},k_{m,z},k_{m,\perp})   \nonumber \\
& & \times{d^3 k_m\over (2\pi)^3 \sqrt{m^2 + k_m^2}} {T_{N^*N}(k_{m,z},k_{m\perp},k_{f,z},k_{f,\perp})\over 4(k_m^2 - k_f^2+m^2_{N*} - m_N^2)},
\label{TNN*}
\end{eqnarray} 
will not require the condition of Eq.(\ref{Vangcond}) with the transition potential having also an imaginary component.  Eq.(\ref{TNN*})
can not be described with any combination of elastic $NN$ interaction potentials that satisfies the angular condition.
The same will be true also  for $\Delta\Delta\to NN$  and   $N_c,N_c\to NN$ transitions.
Thus one concludes that if the $\Gamma_5$ term is not zero  and results in a $k_\perp$ dependence of LF momentum distribution 
then it should originate from a non-nucleonic component in the deuteron.

\section{Predictions and estimate of the possible effects}
\vspace{-0.4cm}
Our calculations predict three new effects, that in probing deuteron structure at very large internal momenta ($\ge m_N$) 
in reaction (\ref{reaction}): (i) the LF momentum distribution should be enhanced compared to $S$- and $D$- wave contributions only;  
(ii) there should be  angular anisotropy  in the LF momentum distribution; (iii) the tensor asymmetry should be significantly different as expected from $S$- and $D$- wave contributions only. 

Observation of all  the above effects will indicate a presence of non-nucleonic components in the deuteron wave function at 
large internal momenta.

\begin{figure}[h]
\begin{center}
\includegraphics[width=7.2cm,height=4.2cm]{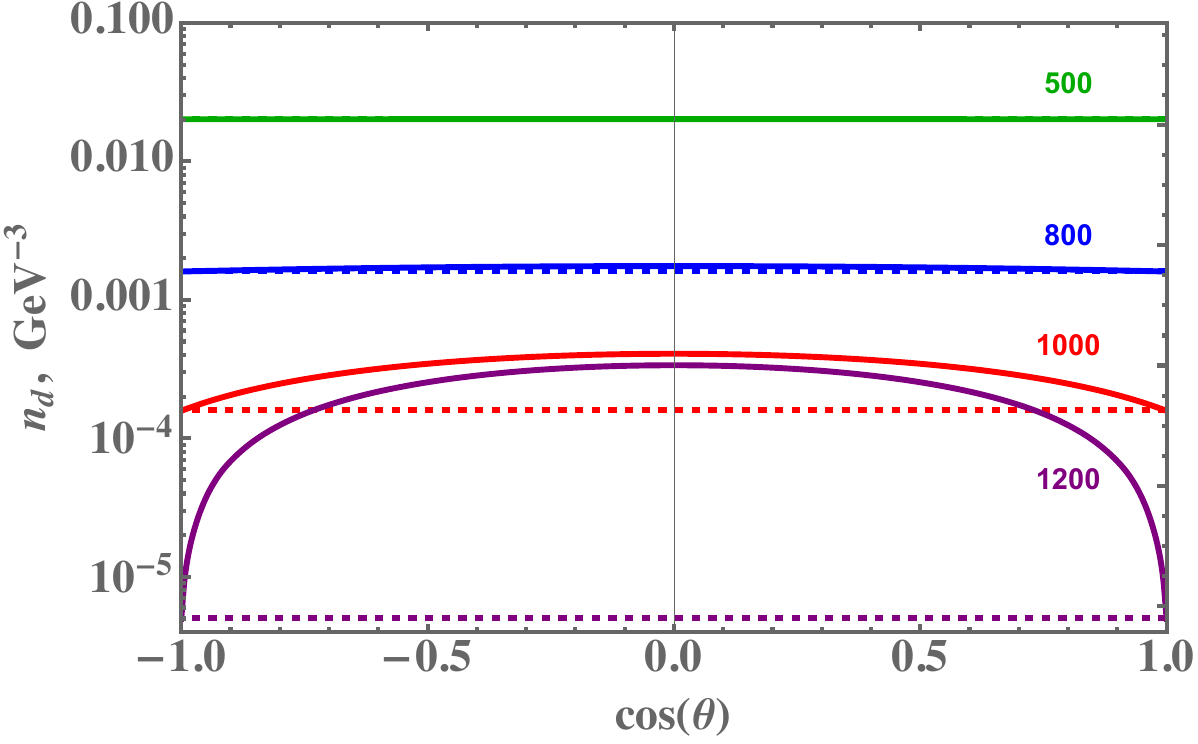}
\includegraphics[width=7.2cm,height=4.2cm]{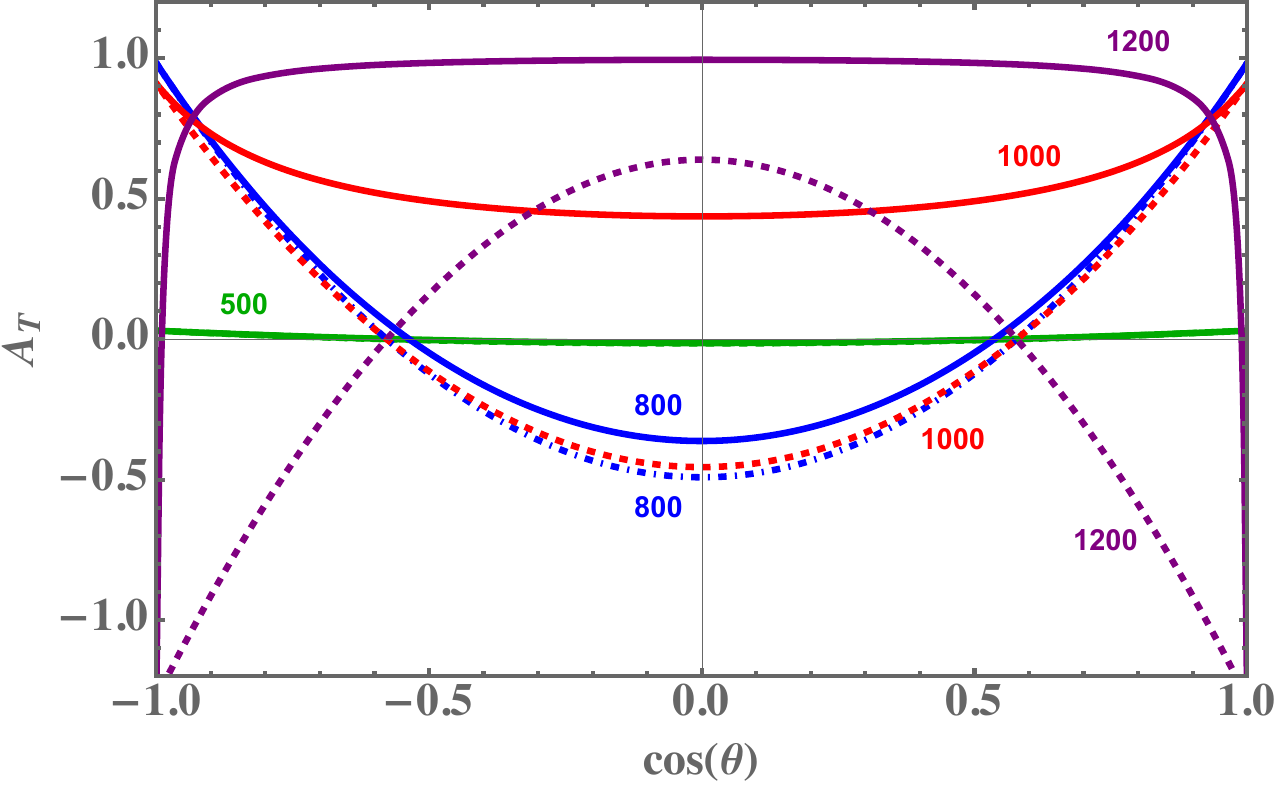}
\vspace{-0.5cm}
\end{center}
\caption{(left panel) LF momentum distribution of the deuteron as a function of $\cos{\theta}$, for different values of $k$.
(right pane). Tensor asymmetry as a function of  $\cos{\theta}$ for different $k$.
 Dashed lines - deuteron with $pn$ component only, solid lines - with $P$-wave like component included.}
\label{mbfigure}
\end{figure}

To give quantitative estimates of the possible effects we evaluate the $\Gamma_5$ vertex function assuming  two color-octet baryon transition to 
the $pn$ system ($N_cN_c\to pn$) through the one-gluon exchange,  parameterizing it in the dipole form ${A\over  (1 + {k^2\over 0.71})^2}$. The parameter $A$ is estimated by assuming 1\% contribution to the total normalization from the $P$ wave.  In Fig.\ref{mbfigure} (right panel) we consider the dependence of the momentum 
distribution of Eq.(\ref{momdist2}) as a function of $\cos{\theta} = {(\alpha-1)E_k\over k}$ for different values of $k$. Notice that if the momentum distribution is 
generated by the $pn$ component only, the angular condition is satisfied, and  no dependence should be observed.

As Fig.\ref{mbfigure} (left panel) shows one may expect  measurable angular dependence  at $k\gtrsim 1$~GeV/c, 
which is consistent with the expectation that  non-nucleonic transition in the deuteron  takes place at $k\gtrsim 800$~MeV/c.  

For tensor polarized deuteron (  Fig.\ref{mbfigure} (right panel)) we estimated the effect using Eq.~(\ref{t20}). 
As the figure shows, in this case,  the presence of a non-nucleonic component will 
be visible   already at $k\approx 800$~MeV/c, resulting in a qualitative difference  
in the asymmetry.

\section{Outlook on experimental verification of the predicted  effects} 
The predictions discussed in the previous section which are related to the existence of  non-nucleonic component in the deuteron 
wave function  can be be verified at CM momenta $k \gtrsim 1$~GeV/c.  These seem 
an incredibly large momenta to be measured in experiment. However, the first such measurement at high $Q^2$ disintegration of the 
deuteron  has already been performed at Jefferson Lab\cite{HallC:2020kdm} reaching $k\sim 1$~GeV/c.  It is intriguing that the 
results of this  measurement qualitatively disagree with  predictions based on conventional deuteron wave functions once $k\gtrsim 800$~MeV/c.
Moreover the data seems to indicate the enhancement of momentum distribution as predicted in our calculations.
 New measurements will significantly improve the quality of the data allowing possible verification of the
second prediction, that is  the existence of angular asymmetry for LF momentum distribution.
What concerns to the tensor asymmetry, it can show a strong sensitivity the non-nucleonic component in the deuteron 
influencing also the repulsive character of bound $pn$ system at very short distancesde.\cite{Sargsian:2024hyx}
Currently there are  significant efforts being made in measuring  high $Q^2$ deuteron electro-disintegration 
processes at Jefferson Lab employing polarized deuteron target\cite{Slifer:2013vma}.

 It is worth mentioning that the analysis of exclusive  deuteron disintegration  experiments will require a careful account for competing nuclear 
effects such as final state  interactions, (FSI) for which there has been significant  theoretical and experimental progress during the last decade\cite{Frankfurt:1996xx,Sargsian:2001ax,Sargsian:2009hf}.
The advantage of high energy scattering is that the eikonal regime is established which makes FSI to be strongly isolated in 
transverse kinematics and be  suppressed in near collinear directions.

If the experiments will  not find the discussed signatures of non-nucleonic components  then they will set a new limit on 
the dominance of the $pn$ component at instantaneous high nuclear densities that corresponds to $\sim 1$~GeV/c internal momentum 
in the deuteron.  However if predictions are confirmed,  they  will motivate theoretical modeling of non-nucleonic components in 
the deuteron, such as $\Delta\Delta$, $N^*N$ or hidden-color $N_cN_c$ that can reproduce the observed results. 
In both cases the results of such 
studies will  advance the understanding of the dynamics of high density nuclear matter and the relevance of the quark-hadron transitions.

 {\bf Acknowledgments:}
This work is supported by the U.S. DOE Office of Nuclear Physics grant DE-FG02-01ER41172.


\begin{thebibliography}{25}


\bibitem{Melnitchouk:1996vp}
W.~Melnitchouk, M.~Sargsian and M.~I.~Strikman,
Z. Phys. A \textbf{359}, 99-109 (1997).

\bibitem{Sargsian:2007gd}
M.~M.~Sargsian,
Nucl. Phys. A \textbf{782}, 199-206 (2007).

\bibitem{Fomin:2010ei}
N.~Fomin (et all) 
Phys. Rev. Lett. \textbf{105}, 212502 (2010).

\bibitem{Freese:2015ebu}
A.~J.~Freese, W.~Cosyn and M.~M.~Sargsian,
Phys. Rev. D \textbf{99}, no.11, 114019 (2019).

\bibitem{Harvey:1980rva}
M.~Harvey,
Nucl. Phys. A \textbf{352}, 326-342 (1981)
doi:10.1016/0375-9474(81)90413-9
\bibitem{Brodsky:1985gt}
S.~J.~Brodsky and C.~R.~Ji,
Phys. Rev. D \textbf{33}, 1406 (1986)
doi:10.1103/PhysRevD.33.1406

\bibitem{Sargsian:2022rmq}
M.~M.~Sargsian and F.~Vera,
Phys. Rev. Lett. \textbf{130}, no.11, 112502 (2023).

\bibitem{Frankfurt:1977vc}
L.~L.~Frankfurt and M.~I.~Strikman,
Nucl. Phys. B \textbf{148}, 107-140 (1979).

\bibitem{Buck:1979ff}
W.~Buck and F.~Gross,
Phys. Rev. D \textbf{20}, 2361 (1979).

\bibitem{Arnold:1980zj}
R.~G.~Arnold, C.~E.~Carlson and F.~Gross,
Phys. Rev. C \textbf{23}, 363 (1981).
 
 \bibitem{Dymarz:1986km}
R.~Dymarz and F.~C.~Khanna,
Phys. Rev. Lett. \textbf{56}, 1448-1451 (1986).

\bibitem{Carbonell:1995yi}
J.~Carbonell and V.~A.~Karmanov,
Nucl. Phys. A \textbf{581}, 625-653 (1995).



\bibitem{Feynman:1973xc}
R.~P.~Feynman, ``Photon-hadron interactions, ''CRC Press (January 1, 1972).
 

\bibitem{Brodsky:1997de}
S.~J.~Brodsky, H.~C.~Pauli and S.~S.~Pinsky,
Phys. Rept. \textbf{301}, 299-486 (1998).


\bibitem{Weinberg:1966jm}
S.~Weinberg,
Phys. Rev. \textbf{150}, 1313-1318 (1966).


\bibitem{Frois:1991wc}L.~L.~Frankfurt and M.~I.~Strikman, in
{\em Modern topics in electron scattering}, edited by 
B.~Frois and I.~Sick, 1991.

\bibitem{HallC:2020kdm}
C.~Yero \textit{et al.} [Hall C],
Phys. Rev. Lett. \textbf{125}, no.26, 262501 (2020).
\bibitem{Sargsian:2024hyx}
M.~M.~Sargsian,
[arXiv:2410.08384 [nucl-th]].

\bibitem{Slifer:2013vma}
K.~Slifer and E.~Long,
PoS \textbf{PSTP2013}, 008 (2013),
[arXiv:1311.4835 [nucl-ex]].
\bibitem{Frankfurt:1996xx}
L.~L.~Frankfurt, M.~M.~Sargsian and M.~I.~Strikman,
Phys. Rev. C \textbf{56}, 1124-1137 (1997).

\bibitem{Sargsian:2001ax}
M.~M.~Sargsian,
Int. J. Mod. Phys. E \textbf{10}, 405-458 (2001).

\bibitem{Sargsian:2009hf}
M.~M.~Sargsian,
Phys. Rev. C \textbf{82}, 014612 (2010).


\end{thebibliography}
\end{document}